%% file: main.tex
\documentclass[times, twoside, watermark]{zHenriquesLab-StyleBioRxiv}

\usepackage[utf8]{inputenc} 
\usepackage[T1]{fontenc}    
\usepackage{url}            
\usepackage{booktabs}       
\usepackage{amsfonts}       
\usepackage{nicefrac}       
\usepackage{microtype}      
\usepackage{graphicx}       
\usepackage{arydshln}
\usepackage{upgreek}
\usepackage{fontawesome}
\usepackage{amsmath}
\usepackage{float}
\usepackage{footnote}
\usepackage{rotating}
\usepackage{multirow}
\usepackage{booktabs}
\usepackage{graphicx} 
\usepackage{amssymb}
\usepackage{relsize}

\leadauthor{Jesus de la Fuente} 
\shorttitle{Sweetwater}

\newcommand{\floor}[1]{\left\lfloor #1 \right\rfloor}

\begin{document}

\title{Sweetwater: An interpretable and adaptive autoencoder for efficient
tissue deconvolution}

\author[1,\dag]{Jesus de la Fuente}
\author[2,\dag]{Naroa Legarra}
\author[3]{Guillermo Serrano}
\author[2,4]{Irene Marín-Goñi}
\author[2]{Aintzane Diaz-Mazkiaran}
\author[2]{Markel Benito Sendin}
\author[2]{Ana García Osta}
\author[5]{Krishna R. Kalari}
\author[6]{Carlos Fernandez-Granda}
\author[1,7,$\ast$]{Idoia Ochoa}
\author[2,7,$\ast$]{Mikel Hernaez}


\affil[1]{TECNUN, University of Navarra, San Sebastián, Spain.}
\affil[2]{CIMA University of Navarra, IdiSNA, Pamplona, Spain.}
\affil[3]{Biological and Environmental Science and Engineering Division, KAUST, Saudi Arabia}
\affil[4]{Department of Molecular Pharmacology and Experimental Therapeutics, Mayo Clinic, USA} 
\affil[5]{Department of Quantitative Health Sciences, Mayo Clinic, Rochester, Minnesota.}
\affil[6]{New York University, New York, United States.}
\affil[7]{Instituto de Ciencia de los Datos e Inteligencia Artificial (DATAI), University of Navarra, Spain}


\maketitle

\begin{abstract}
Single-cell RNA-sequencing (scRNA-seq) stands as a powerful tool for deciphering cellular heterogeneity and exploring gene expression profiles at high resolution. However, its high cost renders it impractical for extensive sample cohorts within routine clinical care, hindering its broader applicability. Hence, many methodologies have recently arised to estimate cell type proportions from bulk RNA-seq samples (known as deconvolution methods). However, they have several limitations: Many depend on selecting a robust scRNA-seq reference dataset, which is often challenging. Secondly, building reliable pseudobulk samples requires determining the optimal number of genes or cells involved in the simulated data generation process, which has not been studied in depth. Moreover, pseudobulk and bulk RNA-seq samples often exhibit distribution shifts. Finally, most modern deconvolution approaches behave as a black box, and the underlying mechanisms of the deconvolution task are still unknown, which can compromise the reliability of the results. In this work, we present Sweetwater, an adaptive and interpretable autoencoder able to efficiently deconvolve bulk RNA-seq and microarray samples leveraging multiple classes of reference data, such as scRNA-seq and single-nuclei RNA-seq. Moreover, it can be trained on a mixture of FACS-sorted FASTQ files, which we newly propose to use as this reduces platform-specific biases and may potentially outperform single-cell-based references. Also, we demonstrate that Sweetwater effectively uncovers biologically meaningful patterns during the training process, increasing the reliability of the results. Sweetwater is available at https://github.com/ubioinformat/Sweetwater, and we anticipate will facilitate and expedite the accurate examination of high-throughput clinical data across diverse applications.
\end {abstract}


\section*{Introduction}



Single-cell RNA sequencing (scRNA-seq) enables the transcriptional profiling of thousands of individual cells, with impressive resolution in cell granularity \cite{lahnemann2020eleven}. However, scRNA-seq still remains a costly technique, vulnerable to noise, and challenging to perform on some tissues, such as the brain, due to cell isolation. Alternatively, the more affordable bulk RNA-seq can provide deeper insights into the sample's gene expression landscape, at the cost of hindering the discovery of cell type specific patterns and functions.

Within the last years, several methodologies have been proposed to estimate cell type proportions from bulk RNA-seq samples by leveraging publicly available scRNA-seq  data. On the one hand, model-driven approaches tackle the deconvolution problem by making data distribution assumptions, such as the gene expression distribution or gene-gene independence (e.g., CIBERSORTx \citep{newman2019determining}, MuSiC \citep{wang2019bulk} or Bayesprism \citep{chu2022cell}). 
These methods rely on a pre-computed reference matrix of gene expression profiles (GEP), often based on specific cell type marker genes. On the other hand, data-driven or deep learning approaches enable interrogation of more genes without relying on pre-computed GEPs or strong statistical assumption on the GEPs, allowing to learn underlying complex gene expression patterns (e.g., Scaden \citep{menden2020deep} or TAPE \citep{chen2022deep}). 

Despite current efforts, deconvolution approaches 
still underperform when generalizing from simulated to real samples, as they are unable to reduce the distribution shifts that arise when training on scRNA-seq and testing on bulk RNA-seq samples. Further, the underlying mechanisms of how these methods obtain cell type proportions given an expression profile is still unknown, limiting the interpretation of the results.

To overcome these limitations, in this work 
i) we present a novel adaptive autoencoder model, termed \textit{Sweetwater}, for estimating cell type fractions, and perform a benchmarking against current deconvolution approaches, evaluating multiple gene sets for estimating cell type proportions; ii) we perform an interpretation analysis on Sweetwater using DeepLIFT \citep{shrikumar2017learning}, and demonstrate that Sweetwater effectively uncovers cell type marker genes throughout the deconvolution task; iii) we then propose a novel way to build pseudobulk samples that allows to minimize technical biases and show that Sweetwater efficiently deconvolves unseen samples, outperforming traditional single-cell based references. Importantly, the proposed method along with the developed tools, are fully available at https://github.com/ubioinformat/Sweetwater.

\section*{Materials and Methods}

\begin{figure}[]
\begin{center}
\includegraphics[width=\linewidth]{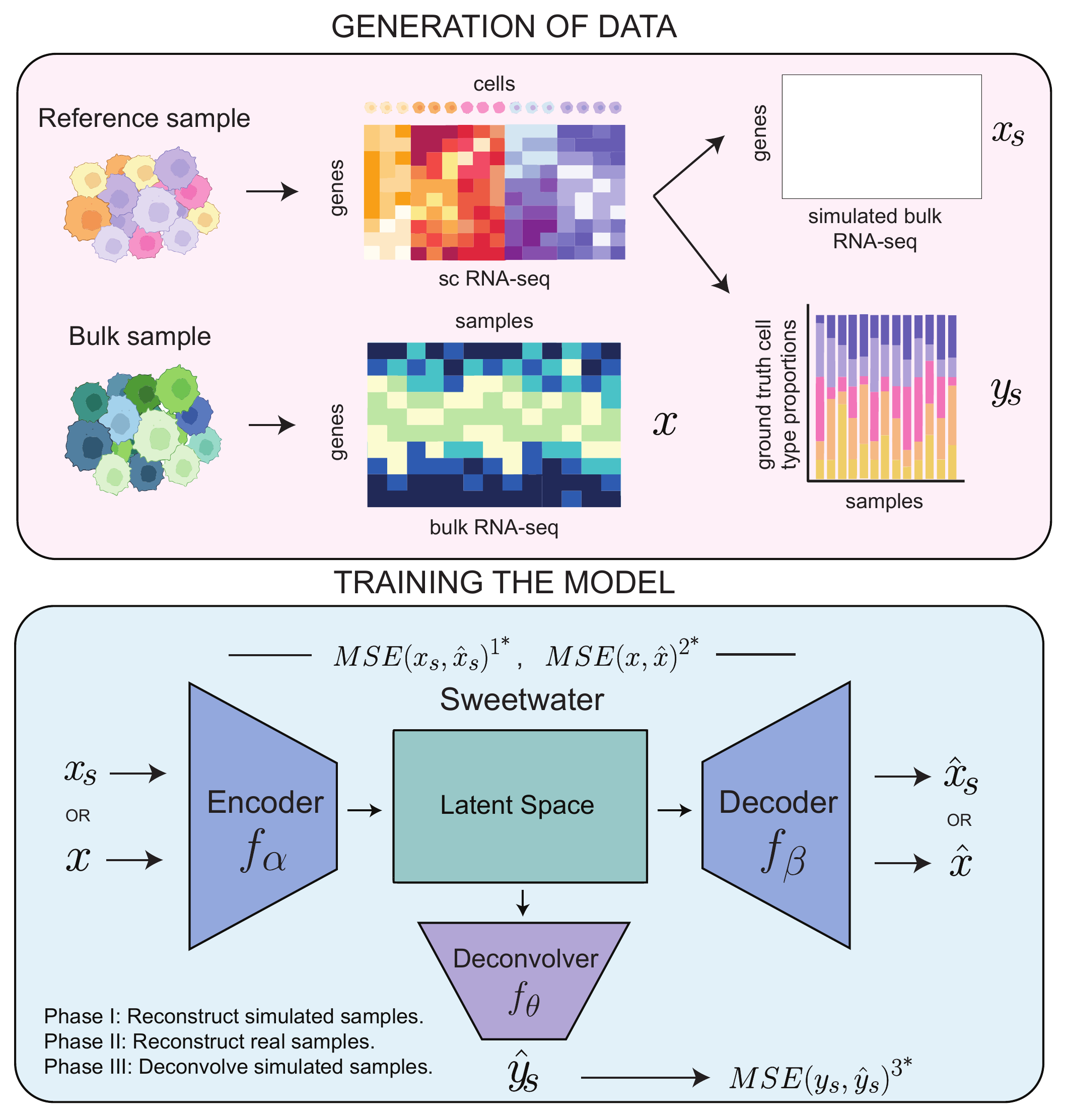}
\end{center}
\caption{\textbf{Sweetwater workflow}. Sweetwater generates pseudobulk samples from a given reference and trains using both simulated and real bulk RNA-seq samples in a three-step process to predict cell type proportions from bulk RNA-seq samples.}
\label{fig:model_overview}
\end{figure}

\subsection*{\textbf{Sweetwater}}

Given an input expression matrix $X \in \mathbb{R}^{G\times S}$, containing the expression of $G$ genes across $S$ samples, the aim of Sweetwater is to deconvolve $X$ into a matrix $Y \in \mathbb{R}^{T\times S}$, whose rows contain the fractions of $T$ cell types for each sample $\mathbf{s}$. For this purpose, $N$ synthetic samples $x_s \in \mathbb{R}^{G}$ and fractions $Y_s \in \mathbb{R}^{G\times T}$ are generated during training from an input scRNA-seq or single-nuclei (sn)RNA-seq matrix $M \in \mathbb{R}^{G\times C}$ containing $C$ cells.  This process requires to randomly subsample cells from $M$ with predefined proportions $y_s$ and aggregating them (via summation) to obtain a synthetic vector $x_s$ with known abundance, which we denote the \textit{pseudobulk} sample. 

\subsubsection*{Model Architecture}


Sweetwater (Figure \ref{fig:model_overview}) is based on a deep autoencoder, composed of three subnetworks: (i) an \textit{encoder} $\mathlarger{\mathlarger{f}}_{\alpha}$: $\mathbb{R}^{G} \rightarrow  \mathbb{R}^{E}$, which contains one hidden layer with one \textit{ReLU} non-linear activation function, (ii) a \textit{decoder} $\mathlarger{\mathlarger{f}}_{\beta}$: $\mathbb{R}^{E} \rightarrow  \mathbb{R}^{G}$, which also consists of one hidden layer and a \textit{ReLU} layer and (iii) a \textit{deconvolver} $\mathlarger{\mathlarger{f}}_{\theta}$: $\mathbb{R}^{E} \rightarrow  \mathbb{R}^{F}$, which contains one hidden layer and a \textit{Softmax} layer, which outputs the deconvolved fractions given a GEP $\textbf{x} \in \mathbb{R}^G$. Here, $E$ denotes the size of the latent space and the \textit{Softmax} layer is chosen so that the deconvolved proportions sum up to one. 

Since Sweetwater is designed to enable deconvolution from multiple-sized GEPs, the proposed architecture presents a dynamic number of neurons depending on the input number of genes.  Specifically, Sweetwater have $G$ neurons at the entry of $\mathlarger{\mathlarger{f}}_{\alpha}$, $\floor{\frac{G}{2}}$ neurons at the hidden layer of $\mathlarger{\mathlarger{f}}_{\alpha}$, and $E = \floor{\frac{G}{4}}$ for the latent space. Decoder $\mathlarger{\mathlarger{f}}_{\beta}$ is symmetric, and $\mathlarger{\mathlarger{f}}_{\theta}$ contains a hidden layer of $\floor{\frac{G}{4}}$ input neurons that yields $T$ cell type fractions after the \textit{Softmax} layer. This ratio (1:$\frac{1}{2}$:$\frac{1}{4}$) presented robust performance with minimum number of trainable parameters across multiple gene sizes (Supp. Fig~\ref{fig:ablation_supp_architecture}).

\subsubsection*{Training Procedure}
Sweetwater is trained in three different phases, where the first two are related to the reconstruction of the synthetic and real expression profiles, and the third phase builds on top of the previous phases to deconvolve the input GEPs into a vector of cell type fractions. 

\begin{figure}[]
\begin{center}
\includegraphics[width=\linewidth]{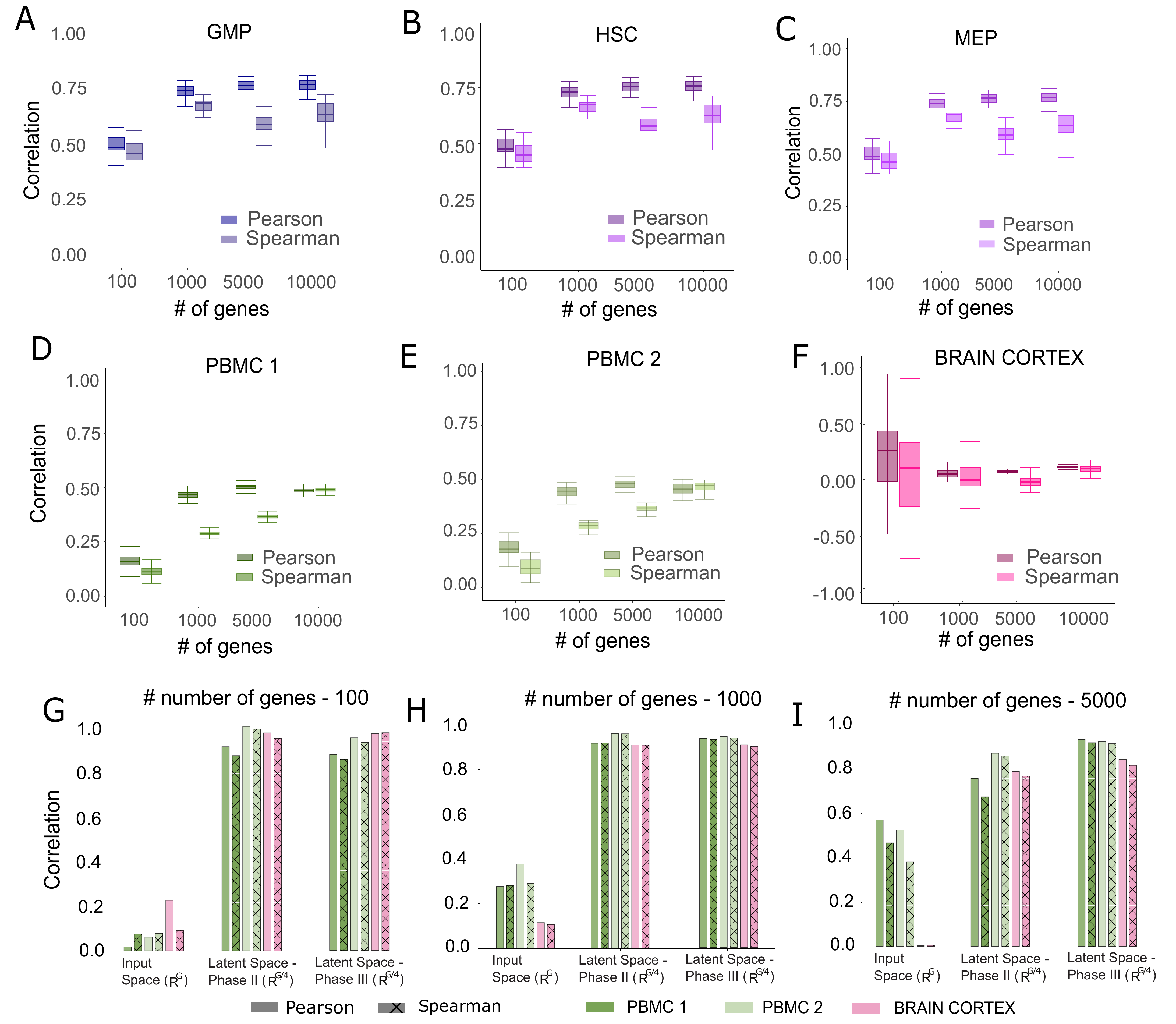}
\end{center}
\caption{\textbf{Correlation on the input gene space and latent space for evaluated datasets}. Pearson and Spearman correlation between real and simulated samples for \textbf{A)} GMP, \textbf{B)} HSC and \textbf{C)} MEP cells and \textbf{D)} Brain Cortex, \textbf{E)} PBMC 1, \textbf{F)} PBMC 2 and datasets. Simulated samples were generated with real ground truth proportions and the respective reference matrix $M$. Also, these correlations were computed at the latent space of Sweetwater after Phase II and Phase II for the evaluated datasets across several group of genes. Note that the left column of G, H and I correspond to this correlation in the original space, which is showed in A-F.}
\label{fig:ablation_study}
\end{figure}

Specifically, in Phase I a low dimensional embedding $\mathbf{z}_s\in\mathbb{R}^E$ is built by minimizing the Mean Squared Error (MSE) between the synthetic samples $\mathbf{x}_s$ and its reconstructed version $\hat{\mathbf{x}}_s=\mathlarger{\mathlarger{f}}_{\beta}(f_{\alpha}(\mathbf{x}_s))$. Then, to minimize platform-specific variation between synthetic and real samples (Figure \ref{fig:ablation_study} A-F), phase II leverages the input GEPs $\mathbf{x}$ to further refine the encoder-decoder networks in a similar way to Phase I. Phase II help transform the samples from a lowly-correlated gene space to a high-correlated latent space (Figure \ref{fig:ablation_study} G-I, middle column). 

Note that even though Phases I and II could be combined into a single Phase, we empirically showed that the proposed training procedure further reduces the reconstruction loss of the synthetic samples $\mathbf{x}_s$ without strictly imposing it (see Supp. Figure \ref{fig:supp_losses_analysis}), and individual Phases help to reach close-to-optimal deconvolution performance (see Phase III) in a faster and more stable way than a combined Phase (results not shown). Note that during Phase I and II the weights of $\mathlarger{\mathlarger{f}}_{\theta}$ remained fixed.

Finally, Phase III is trained to deconvolve the synthetic samples $\mathbf{x}_s$ into the known proportions $\mathbf{y}_s$ by leveraging the acquired knowledge and gradients from preceding phases. Here, MSE is also used between the deconvolved $\hat{\mathbf{y}}_s$ proportions and true ones $\mathbf{y}_s$ as it penalizes incorrect predictions more severely. Hence, during Phase III $\mathlarger{\mathlarger{f}}_{\beta}$ remains fixed, and $\mathlarger{\mathlarger{f}}_{\alpha}$ and $\mathlarger{\mathlarger{f}}_{\theta}$ are updated. It is worth mentioning that here the encoder  $\mathlarger{\mathlarger{f}}_{\alpha}$ does not remain fixed, as we empirically showed that this Phase also helps to increase the correlation in the latent space between synthetic and real samples as the number of input genes increases, allowing a better generalization capabilities (Figure \ref{fig:ablation_study} G-I, right column). See Supp. Note \ref{notes:details} for further details on training procedure. 

\subsection*{\textbf{Datasets and preprocessing}}

\subsubsection*{Human Bone Marrow}

For young healthy human bone marrow samples, we employed a scRNA-seq matrix and a bulk RNA-seq dataset of isolated hematopoietic stem cells (HSC), megakaryocyte-erythroid progenitors (MEP), and granulocyte-monocyte progenitors (GMP). The scRNA-seq matrix was constructed from the integration of 5 different samples from  \citep{ainciburu2023uncovering} (GSE180298), using Seurat v4.3. Additionally, the bulk RNA-seq dataset was generated from 9 samples of fluorescence-activated cell sorting (FACS) isolated HSC, MEP, and GMP cell types and was quantified using the default parameters of Kallisto v0.46. FACS, a specialized form of flow cytometry, focuses on isolating targeted cell populations while also providing estimates of the number of cells isolated.

\subsubsection*{Human Peripheral Blood Mononuclear Cells}


For human peripheral blood mononuclear cell (PBMC) tissue we used one scRNA-seq matrix from the 10X Genomics data website anotated using Seurat (version 4.3.0.1) \citep{butler2018integrating} and two microarray datasets (GSE107990 and GSE65133) with ground-truth proportions quantified by FACS. Proportions for every study were obtained from the Supplementary Table 6 of \citep{monaco2019rna} and were normalized to sum to one for microarray GSE107990. For microarray GSE65133, proportions were originally obtained in relation to the PBMC fraction of blood, hence a new column of \textit{unknown} cell types was added to every sample, to also enforce them to sum up to one. 
\subsubsection*{Human Brain Cortex}

For the healthy human brain cortex we utilized one snRNA-seq sample (GSM4982106) from the GEO database (GSE163577) \citep{yang2022human} and one bulk RNA-seq dataset from the Religious Orders Study (ROS) and Rush Memory and Aging Project (MAP). Cell type proportions of the bulk RNA-seq were calculated by image analysis of a parallel IHC dataset (\href{https://github.com/ellispatrick/CortexCellDeconv}{link}).  

\subsubsection*{Input data preprocessing}

To maintain consistency, common genes across real and synthetic samples were selected. Also, several variance cutoff thresholds have been applied to remove non-informative genes, ranging from the top 100, 1k, 5k up to 10k genes. Finally, as proposed in \citep{menden2020deep}, we performed a two-step preprocessing of the input GEP vector, first by accounting for heteroscedasticity by applying a non-linear transformation to the input vector, and then scaling to the range [0,1]. We refer to \citep{menden2020deep} for a detailed explanation. This data processing workflow was also applied to the developed baseline model.

\subsubsection*{Pseudobulk sample generation from scRNA-seq data}
In order to train Sweetwater, a substantial amount of GEPs with known cell type fractions are needed. Hence, scRNA-seq or snRNA-seq matrices were used to generate synthetic GEPs with known cell type proportions. Note that we assume the sc/snRNA-seq matrix $M$ to be annotated, i.e., each cell has an associated cell type.  In order to generate $N$ ($\mathbf{x}_s$,$\mathbf{y}_s$) tuples, we first define a parameter $v$, which denotes the number of cells to be subsampled for each sample. For each sample to be generated, we define $\mathbf{y}_s$, containing the proportion of $T$ cell types in that sample. Being $y_s[t]$ the $t^{th}$ proportion in the $y_s$ array, for each cell type $t$, we randomly subsample $\floor{y_s[t]*n}$ cells from the submatrix $M_t \in M$, composed of all the cells belonging to the cell type $t$. Finally we sum the resultant cells ($\mathbb{R}^{v\times G}\rightarrow \mathbb{R}^{G}$) yielding the \textit{pseudobulk} sample $\mathbf{x}_s$. We repeated this process $N$ times, varying the number of used cells $v$ from 100 up to 10k to generate a more heterogeneous simulated dataset. See Supp. Note \ref{notes:details} for further details on $y_s$ generation.

\subsubsection*{Pseudobulk sample generation from FASTQ files}
Sweetwater is able to leverage a mixture of raw sequencing files (FASTQ files) of FACS isolated cell types for training. To generate these samples, FASTQ files should correspond to GEPs from single isolated cell types. We denote $Q$ the set of FACS-sorted FASTQ files. In order to generate a synthetic sample $x^{Fq}_{s}$, for each cell type $t \in T$, we select the corresponding file $q_t \in Q$ associated to cell type $t$. Note that if more than one FASTQ file is available for cell type $t$ due to for example, several donors, all these files are concatenated into a unique FASTQ file $q_t$. Then we subsample lines in groups of 4 (because of the FASTQ structure) proportionally to $y_{s}[t]$. All the subsampled lines are randomly pasted into a new FASTQ file, which is then aligned and quantified using Kallisto v0.46 (single end, 200 fragment length and 20 standard deviation), to generate the resulting $x^{Fq}_{s}$.

\subsection*{\textbf{Interpretability of Sweetwater with DeepLIFT}}

DeepLIFT \citep{shrikumar2017learning} assigns contribution scores to each input feature, indicating how much each feature contributes to the model's output. DeepLIFT operates by comparing the activation of a particular neuron or output in the model to a reference activation, which is usually chosen to represent some baseline or expected behavior. For the performed tasks, we used a matrix of zeroes as the DeepLIFT reference, and computed each gene contribution in $\mathlarger{\mathlarger{f}}_{\alpha}$ input layer to every $\mathlarger{\mathlarger{f}}_{\theta}$'s output neuron, representing the cell type proportions of the deconvolved sample.

\subsection*{\textbf{Functional enrichment analysis}}
To perform functional enrichment analysis on biological process (BP) of Gene Ontology sets we used  clusterProfiler \citep{clusterProfiler} and org.Hs.eg.db v3.14\citep{OrgDb} as annotation data. We input DeepLIFT-ranked top-500 genes for each cell type proportion group after name conversion to ENSEMBL IDs using AnnotationDbi v1.56 \citep{AnnotationDbi}.

\subsection*{\textbf{Evaluation metrics}}
For evaluating the inferred cell type fractions, we used Mean Absolute Error (MAE) \citep{chen2022deep,menden2020deep} and Lin's Concordance Correlation Coefficient (CCC) \citep{lawrence1989concordance}. These metrics measure not only the absolute deviation from the ground-truth, but also the correlation of each cell type, allowing for a better understanding of the predicted values. The MAE and CCC are computed as
\begin{equation*}
    \begin{aligned}
        & MAE(y,\hat{y}) = \frac{\sum_{ij} |y_{ij} - \hat{y}_{ij}| }{n \times C_t};\\
        & CCC(y, \hat{y}) = \frac{2 \times cov(y,\hat{y})}{\sigma^{2}_{y} + \sigma^{2}_{\hat{y}} + (\mu_{y}-\mu_{\hat{y}})},
    \end{aligned}
\end{equation*}
where $cov(y,\hat{y})$ stands for the covariance between the predicted and real proportions and $\mu$ and $\sigma$  represent the mean and standard deviation, respectively.

\section*{Results}

\subsection*{\textbf{Sweetwater performs efficient deconvolution on both simulated and real bulk samples}}

We evaluated Sweetwater against state-of-the-art deconvolution approaches such as Scaden \citep{menden2020deep} and TAPE \citep{chen2022deep}, the two most known and recently published data-driven approaches, and CIBERSORTx \citep{newman2019determining} and Bayesprism \citep{chu2022cell}, two of the latest and most cited model-driven methodologies. Additionally, we considered one data-driven baseline model, denoted `Baseline XGB', based on a gradient boosting model. This benchmarking was performed for the aforementioned datasets: PBMC 1, PBMC 2 and Brain Cortex. As most of the evaluated approaches use a reference matrix to obtain the deconvolved tissue proportions, we first benchmark the deconvolution approaches on simulated bulk data. This analysis allows to benchmark them without considering batch effects or train/test distribution shifts that may arise when using real bulk samples. Therefore, we can estimate the degree to which these models are effectively capturing the intrinsic structure of the reference matrix. We generated simulated data (Supp. Figure \ref{fig:supp_test_matrices}) following each method's guidelines, and evaluated them using the top 100, 1k, 5k and 10k most variant genes.

\begin{figure*}[]
\begin{center}
\includegraphics[width=\linewidth]{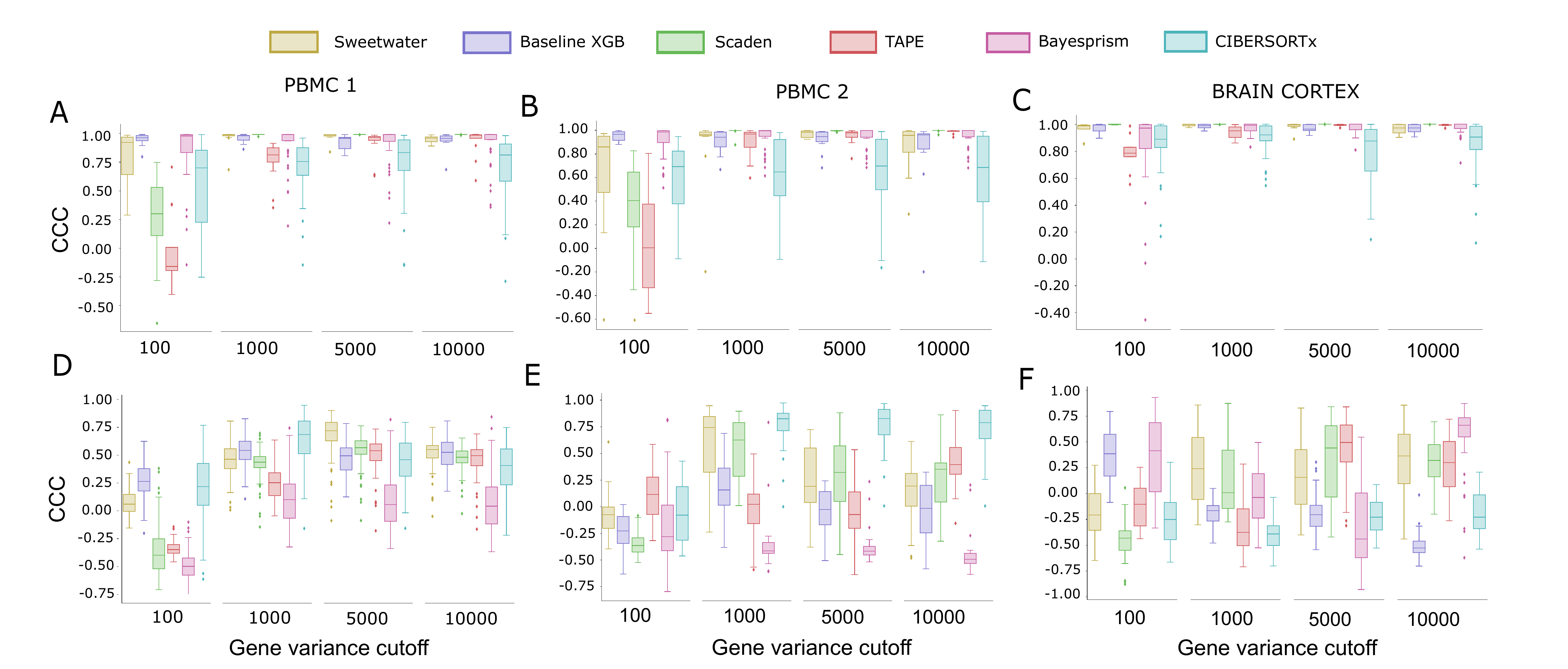}
\end{center}
\caption{\textbf{Deconvolution approaches benchmarking}. \textbf{A-C}. CCC of deconvolved against ground truth proportions for simulated bulk in PBMC 1 (\textbf{A}), PBMC 2 (\textbf{B}), and Brain Cortex datasets (\textbf{C}), and for real bulk in PBMC 1 (\textbf{D}), PBMC 2 (\textbf{E}) and Brain Cortex datasets (\textbf{F}), accross different top variant gene cutoffs. Results for different state-of-the-art models are showed in the same order as the legend. The parameters used for the XGB Baseline model: n\_estimators=1000, max\_depth=10, eta=0.1, subsample=0.7, colsample\_bytree=0.8. We used 16 cores for each deconvolution task. }

\label{fig:model_comparison}
\end{figure*}

Figure~\ref{fig:model_comparison} shows the deconvolution benchmarking for PBMC 1, PBMC 2 and Brain Cortex datasets. Note that even though PBMC 1 and PBMC 2 share the same reference matrix, it is used differently in both datasets, as the associated microarrays have different number of genes and cell types, hence the resulting simulated samples differ. For every evaluated dataset, we notice that most of the state-of-the-art deconvolution models were able to achieve a high correlation on pseudobulk samples (Figure~\ref{fig:model_comparison} A-C right) between the predicted test and the ground-truth cell type proportions. In fact, even a small number (100 to 1k) of the top variant genes is enough to obtain almost perfect correlation on the test set. It is worth mentioning that CIBERSORTx obtained the lowest CCC on average, which could be expected given the discrepancies seen between pseudobulk and bulk samples and the fact that it has not been designed to deconvolve such type of data.

When these models were evaluated on real bulk RNA-seq (Brain Cortex) and microarray (PBMC 1 and 2) samples, a notable decrease on test correlation across models and group of genes can be appreciated (Figure~\ref{fig:model_comparison} A-C left). 

Despite of this, Sweetwater was able to obtain best CCC performance on PBMC 1 when a large enough number of genes was selected, and above average with low number of available genes (100 and 1000). CIBERSORTx showed best results on PBMC 2 followed by Scaden an Sweetwater, which again yield above average performance on Brain Cortex tissue when an elevated number of genes was selected, proving consistency against more unstable models such as TAPE or Bayesprism, which presented best (Bayesprism, 10000 genes) and worse (Bayesprism, 5000 genes) results depending on the number of genes selected. Also, although some approaches obtained almost perfect test CCC with very low variance when evaluated on simulated samples, such as Scaden or Bayesprism, this is not translated to real samples, where the variance largely increased and the CCC notably dropped. This may imply a significant distribution shift between simulated and actual bulk samples, a challenge that many models encounter when trying to generalize effectively. We believe the latent space built by Sweetwater helped to achieve this stability, reducing the aforementioned distribution shift. Finally, it should be emphasized that the baseline model XGB, even though obtaining a very good performance on simulated samples, suffered from the aforementioned distribution shift, and failed to effectively deconvolve real samples.

Throughout the deconvolution task, we measured the time consumption across genes and datasets (Supp. Figure \ref{fig:supp_time_comparison}). Among the evaluated models, XGB and CIBERSORTx reported the lowest time consumption, closely followed by our methodology Sweetwater. Both CIBERSORTx and XGB models reported values ranging from $\approx$ 90 seconds up to 15 minutes when going up in the number of used genes. On the other hand, Scaden was the worst performing model on time consumption, reporting an unusual increment when going down in the number of used genes for deconvolution, with up to 3 hours/run. This may be due to the incapability of Scaden to converge when working with low number of genes, as reported on the test CCC for simulated data (top 100 variant genes). Sweetwater also suffered from an increase in time consumption when only using the top 100 variant genes, but reported close to state-of-the-art values on the remaining group of genes. Also, it is worth mentioning that one of the best time performers, CIBERSORTx, scales really poorly compared to other machine learning approaches such as XGB or Sweetwater, as it runs individually for each of the deconvolved samples.

These findings highlight that the proposed method Sweetwater is able to maintain an stable and above-average high correlation, along with a low mean absolute error (Supp. Figure~\ref{fig:supp_benchmarking_MAE}) on both simulated and real microarray and bulk RNA-seq samples, in a reasonably fast manner. Also, these results suggest that there is still gap for improvement when generalizing from simulated to real samples, which may require minimizing existing distribution shifts, related to different sources such as technical bias, or building more reliable simulated data, exploring non-linear cells aggregation. See Supplementary Note \ref{notes:bp_analysis} for a detailed analysis on discrepancies between real simulated data.

\subsection*{\textbf{Sweetwater learns biologically meaningful patterns during the training process}}
One major drawback of current state-of-the-art deconvolution approaches is their lack of interpretability. This is compounded by the complexity of current methods, as many employ deep neural networks often perceived as black boxes. Here, we dissect the underlying relationships Sweetwater learns during training via DeepLIFT \citep{shrikumar2017learning}, a method that infers the contribution of each input feature to each model's output, to uncover what genes are the most important for the deconvolution process. For this analysis, we computed the contribution score of each input gene to every cell type proportion, by applying DeepLIFT on the weights Sweetwater learned when deconvolving PBMC 1 and Brain Cortex datasets. See methods for further details on how DeepLIFT works.

\begin{figure}[]
\begin{center}
\includegraphics[width=\linewidth]{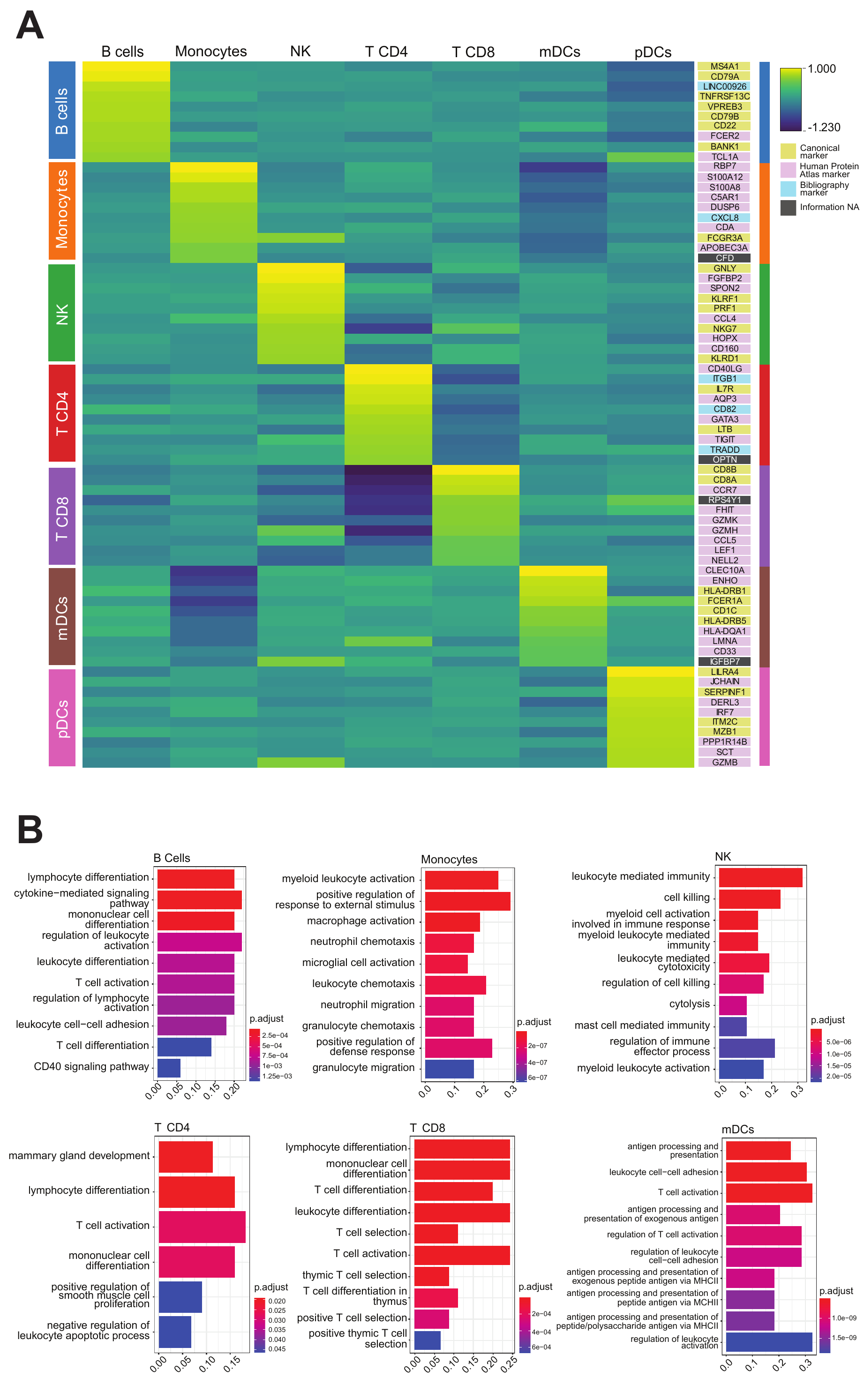}
\end{center}
\caption{\textbf{Sweetwater interpretation with DeepLIFT}. A) Heatmap showing the top 10 genes for each cell type according to DeepLIFT score, when training Sweetwater on PBMC 1 dataset. B) Barplots showing top enriched GO terms (BP) for the top 500 genes for each cell type according to DeepLIFT score, when training Sweetwater on PBMC 1 dataset.}
\label{fig:model_interpretation}
\end{figure}

We first examined the top 10 genes with the highest DeepLIFT scores for each cell type, revealing that most of them played an important role as marker genes of that cell types. This finding strongly suggests that Sweetwater employs biological information as a fundamental driver for the deconvolution processes. In the PBMC 1 dataset, most top genes were established marker genes for their corresponding cell types, while others such as TRADD \citep{pobezinskaya2011adaptor}, ITGB1 \citep{nicolet2021cd29} or CD82 \citep{shibagaki1999overexpression} in CD4\textsuperscript{+} T lymphocytes (CD4 T),  CXCL8 in monocytes \citep{walz1987purification} and LINC00926 in B-cells  \citep{li202111p}, have been described to have very important functional roles (Figure \ref{fig:model_interpretation}-A). Indeed, recent studies have identified LINC00926 as a long non-coding gene uniquely expressed in B-cells. Similarly, the significance of ITGB1 in T CD4 cells has been highlighted in recent findings, specifically for its important role in a distinct subtype of cytotoxic T CD4 cells. We conducted a differential expression analysis on these genes within the reference scRNA-seq matrix across the different cell types, revealing a notable trend: genes with higher DeepLIFT scores relative to other cell types also exhibit higher log fold changes (Supp. Figure~\ref{fig:DE_deeplift}). These findings provide further confirmation that marker genes play a crucial role in the deconvolution process.

To reinforce our hypothesis that our method is indeed learning biologically meaningful patterns, we expanded the list of DeepLIFT-ranked genes up to the top 500 for each cell type and performed functional enrichment analysis on Gene Ontology (GO) biological process annotations. We expected top enriched pathways in each cell type proportion category to agree well with the known canonical cell functions. 

Indeed, we found concordant results for most cell types (Figure~\ref{fig:model_interpretation}-B). Natural killer (NK) cell is an immune cell with cytolytic properties involved in innate immune response \citep{NKcells} accordingly top enriched pathways are related with cytotoxicity, cytolysis and regulation of immune response. Another example: almost all top enriched GO terms for CD8\textsuperscript{+} T cells were related with specific T-cell selection, differentiation or activation processes. Similar matching results were found also for myeloid dendritic cells (mDC), B-cells, monocytes and T CD4, however plasmacytoid dendritic cells (pDC) did not show any significant enriched term.

All together, this enables the interpretation of Sweetwater's learning procedure in the context of the deconvolution task, validating the training effectiveness. Refer to Supp. Note \ref{notes:deeplift} for DeepLIFT analysis on the Brain Cortex dataset.

\subsection*{\textbf{Sweetwater boosts deconvolution performance through isolated-FASTQ files mixture}} 

\begin{figure}[]
\begin{center}
\includegraphics[width=\linewidth]{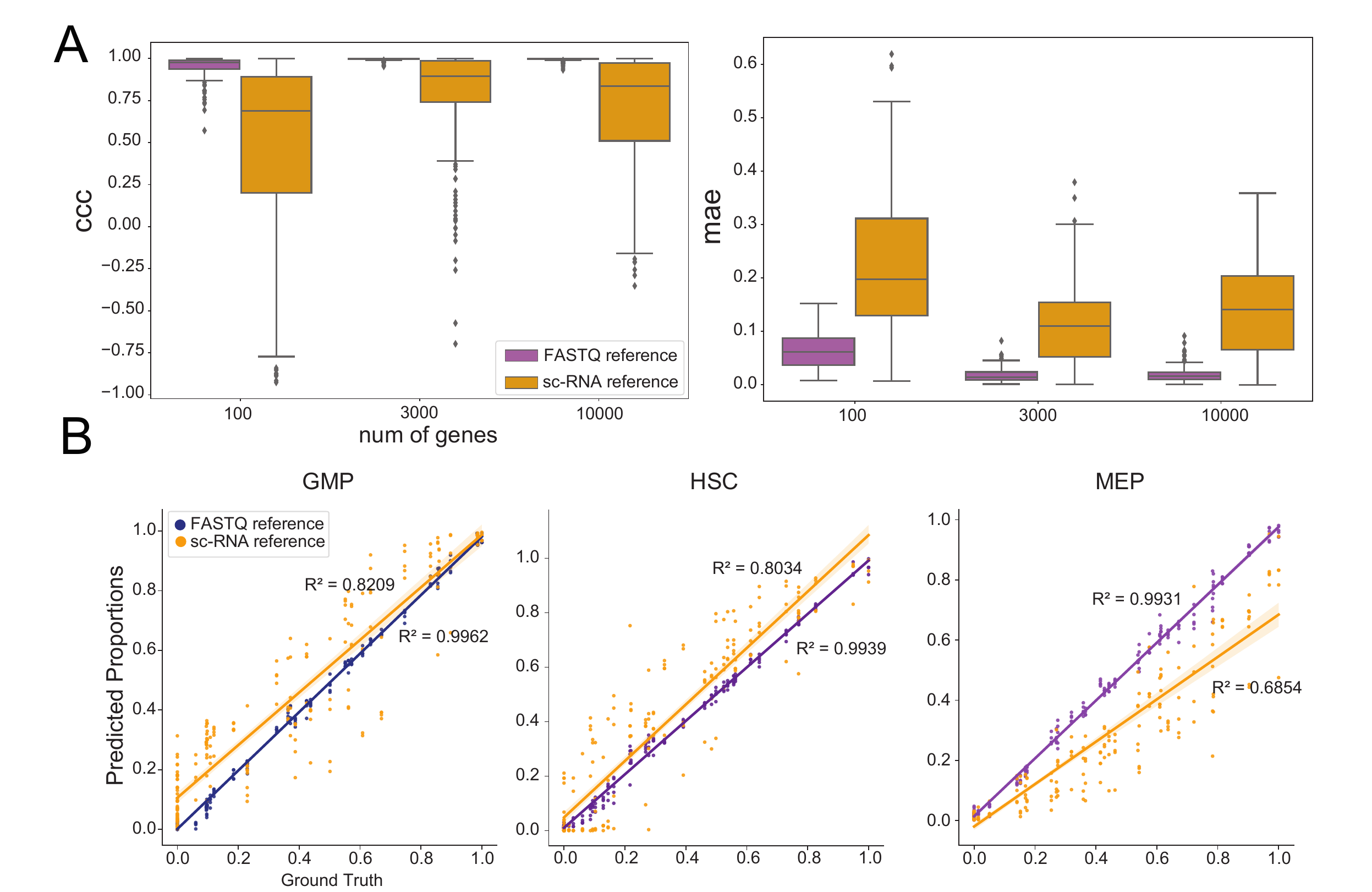}
\end{center}
\caption{\textbf{Leveraging FASTQ files to deconvolve simulated human bone marrow bulk RNA-seq samples}. \textbf{A}. CCC and MAE of both models (scRNA-seq and mixed FASTQ files). Several groups of most variant genes have been selected. \textbf{B}. Scatter plots showing, for each cell type comprising our dataset, the real vs. the predicted proportions, using the top 3000 most variant genes. Linear regression has been fitted for each of the models, and the $R^{2}$ coefficient has been computed.}
\label{fig:fastq_model}
\end{figure}

While scRNA-seq data allows to generate pseudobulk samples for data-driven approaches, this task is still challenging for several reasons. Indeed, a similar reference to the bulk samples aimed to be deconvolved is not always available, particularly when addressing disease-related phenotypes. Additionally, even if a suitable reference is found, an elevated number of cells for each cell type is often demanded. Hence, dealing with low abundance cell types may incur in low variance samples, hindering to capture bulk RNA-seq sample complexity. Finally, scRNA-seq or snRNA-seq technologies introduce strong technical bias, which can confound the model and create a distribution shift between train and test data.

Here, we propose a novel way to generate pseudobulk samples that aims to address these issues. Supp. Figure~\ref{fig:supp_fastq} depicts the data preparation process for the deconvolution of the Bone Marrow dataset. On one side, a reference scRNA-seq matrix, built by integration of 5 different donors, was used to generate simulated bulk samples in the traditional way. On the other side, we propose a new way of constructing the simulated samples used for training by subsampling reads from bulkRNA-seq FASTQ files of isolated cell types, thus minimizing the distribution shift between simulated and real data. The employed dataset is composed of bulkRNA-seq FASTQ files of FACS isolated GMP, HSC and MEP cell types in each of the 9 donor samples. 7 samples of each cell type were selected for training and 2 were left out for testing. We randomly subsampled a percentage of the reads from each cell type FASTQ files and concatenated them into a new file, prior to alingment and quantification with Kallisto. Following this procedure, we generated 1000 and 200 samples for train and test folds, respectively, including a wide range of different cell type proportions. Therefore, two different training sets were built, one using scRNA-seq matrix in the traditional way already described, and another following the novel proposed way. 

We evaluated both models across three different groups of most variant genes (100, 3000 and 10000), and showed that when FASTQ reference is used, Sweetwater is able to effectively deconvolve test samples obtaining almost perfect correlation and minimum error (Figure~\ref{fig:fastq_model}-A). From the genes' group that best minimized the MAE for both FASTQ and scRNA-seq references ($ngenes=3000$), we compared ground truth against predicted proportions. Sweetwater minimizes better the differences between real and predicted proportions when FASTQ files are used as reference (Figure~\ref{fig:fastq_model}-B).

These results suggest that the proposed way of leveraging FASTQ files has enabled Sweetwater to successfully deconvolve unseen samples while minimizing technical bias and distribution shifts, outperforming traditional ways of generating simulated data such as scRNA-seq pseudobulk samples.

\section*{Discussion and Conclusion}


In this work we presented a novel adaptive autoencoder that obtained state-of-the-art results across different groups of genes and datasets showing that its learning process involves uncovering biological meaningful patterns, such as the identification of cell type marker genes. 


We first benchmarked Sweetwater against multiple state-of-the-art deconvolution models, consistently achieving top CCC scores across most datasets, outperforming most methods in time consumption. Also, it was shown that an inherent gap still persists in deconvolving pseudobulk and bulk samples that current methods cannot address. This underscores the need for further work in building reliable simulated samples that minimize the distribution shift with real data.

We then employed explainability models to scrutinize the intrinsic relationships Sweetwater learns during the deconvolution training process, allowing us to unveil the contribution of each input feature to the model's output. We found that top contributing genes are not only well-established canonical markers but also recently identified genes with lower expression levels yet of high functional relevance. Additionally, Gene Ontology analysis demonstrated enriched pathways aligned well with known canonical cell functions, thus validating the model's ability to discern authentic and well-defined cell-specific biological functions.

Moreover, we presented, to the best of our knowledge, the first technique to build pseudobulk samples leveraging bulk FASTQ files and proved that it can potentially outperform traditional scRNA-seq references. We hypothesized that this may be transferable to non-synthetic bulk samples, potentially positioning this technique as a useful tool for reducing the limitations of scRNA-seq based references.

Overall, we envision that this work will not only provide a novel efficient deconvolution technology, but also an innovative as well as advantageous way to leverage sequencing data, such as mixed bulk FASTQ files, allowing to perform deconvolution on environments where scRNA-seq may not be available. Moreover, the interpretability feature of Sweetwater will enable to uncover biological patterns along the learning process, and thus become a valuable tool for measuring the reliability of the results. We have made Sweetwater available as a Python package at https://github.com/ubioinformat/Sweetwater.

\section*{Competing interests}
No competing interest is declared.

\section*{Author contributions statement}
\textbf{J.F}: Conceptualization, software, visualization, methodology, writing. \textbf{N.L}: Conceptualization, software, visualization, methodology, writing. \textbf{G.S}: Conceptualization, visualization. \textbf{I.M}: Visualization, writing. \textbf{A.D}: Data generation (Bone Marrow dataset). \textbf{M.B}: Benchmarking process (CIBERSORTx). \textbf{R.K}: Conceptualization. \textbf{C.F}: Conceptualization, methodology. \textbf{I.O}: Conceptualization, supervision, methodology, writing. \textbf{M.H}: Conceptualization, supervision, methodology, writing. Authors have stated that there is no competing interests.

\section*{Acknowledgments}
This work was supported by the following grants: RYC2021- 033127-I. ISCIII ``AC23\_2/00016". MCIN/AEI TED2021 - 131300B - I00. Fulbright Predoctoral Research Program [PS00342367], a Fundacion Ramon Areces predoctoral grant, MCIN/AEI RYC2019-028578-I, Gipuzkoa Fellows (2022-FELL-000003-01), and MCIN/AEI (PID2021-126718OA-I00).

\clearpage
\bibliography{references}

\newpage
\onecolumn
\input{supp_notes}

\newpage
\input{supp_figures}


\end{document}

%% file: supp_notes.tex
\setcounter{section}{0}
\setcounter{page}{1} 
\renewcommand{\thesection}{S\arabic{section}}



\section{Additional details of Sweetwater}\label{notes:details}

\subsection*{Data split} Through the work, we used different sample sizes $N$ for training and evaluating several state-of-the-art deconvolution approaches. For Sweetwater and our designed XGB baseline, we used a train/test split of 0.8:0.2, that was performed with the scikit-learn function ``train\_test\_split''. For the rest of the evaluated methodologies, we followed the recommended split used in the original manuscript. This split was applied during Phase I to avoid overfitting and allow generalization to unseen samples, hence the test MSE loss is monitored using an EarlyStopping while $f_{\alpha}$ and $f_{\beta}$ weights are updated to allow the reconstruction of the simulated samples $X_s$. In a similar manner, during Phase II a train/test is applied over the bulkRNA-seq or microarray samples $X$ in a 0.8:0.2 ratio and test MSE loss is also monitored using an EarlyStopping. Phase III use the same split and data described in Phase I, incorporating the proportions of the simulated samples.\\

\subsection*{Training procedure} During the training procedure, multiple EarlyStoppings were implemented to avoid overfitting. This approach works as follows. A \textit{patience} variable $p$ is initially defined ($ p = p_0$), and a loss function, from the test fold in our case, is monitored during the training stage, and its checked at the end of every epoch. Every time we encounter a minimum in our monitored loss $l_m$, we stored it, and for every epoch in which the current loss $l_c$ is higher to the minimum encountered ($l_c > l_m$), we update our patience variable $p = p-1$. If we find a new minimum, we update $l_m$ and reset $p$, i.e., $p = p_0$. Training will stop when $p = 0$. \\We applied an EarlyStopping for each training Phase, with a patience of 10 for Phase I and II, and a patience of 50 for Phase III. Along the three training Phases, an Adam optimizer \cite{kingma2014adam} with a learning rate of $1e^{-5}$ is used to optimize the network parameters. Even though EarlyStoppings were implemented, we usually found the training of Phase III stable after 30000 samples. Hence, we defined the number of epochs $e$ as $e = \floor{30000/(N_{train}/bs)}$, where $N_{train}$ is the number of samples used for the training fold and $bs_s$ is the batch size of simulated samples, that is set to 256. During Phase II, and due to the low number of samples usually available in bulkRNA-seq and microarray, the batch size $bs$ for real samples is set to 1. Same epochs are used for Phase I and Phase II, although usually require low number of iterations, hence the low patience value of the EarlyStopping.\\

\subsection*{Simulation of pseudobulk data} Already published methodologies \cite{menden2020deep,chen2022deep} used to generate simulated RNA-seq samples following a two-fold approach, in which they distinguished among "normal" samples and "sparse" samples. These approaches first used the \textit{dirichlet()} function from the \textit{numpy.random} package to generate the "normal" samples, as the distribution is desired to be uniform (not to introduce cell type specific biases) and summing up to 1, to fulfill the percentages requirements. Then, to generate the random samples, some specific cell types would be randomly chosen, set to 0, and those samples will be normalized to sum up to 1. \\Even though this approach will force the model to learn to identify non-present cell types, a problem arises when dealing with high number of them. As the uniform distribution's average is $1/T$, having large amount of cell types may avoid samples to have high percentage of them, preventing the model to learn to identify highly present cell types within a RNA-seq. In order to address this, Sweetwater generates every combination of cell types,  i.e., $\sum_i^{T} \frac{T!}{(T-i)!i!}$, and sampled from the Dirichlet's distribution to have samples with unit sum, allowing highly-sparse samples to be present in the simulated RNA-seq dataset, increasing the space of cell type proportions possibilities the model can learn.

\newpage
\section{Evaluation of technical disparities between bulk and pseudobulk samples}\label{notes:bp_analysis}
\subsection*{Differential expression and correlation analysis between bulk and pseudobulk RNA-seq samples}

We generated pseudobulk samples by subsampling $v = \{10, 100, 1k, 5k, 10k\}$ cells from the corresponding scRNA-seq matrices, preserving proportions consistent with each bulk RNA-seq sample. We performed unpaired differential gene expression analysis between pseudobulk and bulk samples using DESeq2 \citep{love2014moderated} v1.38. Genes with an adjusted p-value $\leq$ 0.05 and a log-fold change (LFC) $\geq$ 1 were selected as differentially expressed, and the resulting percentage was calculated relative to the total number of considered genes. For the correlation analysis, we calculated paired Pearson and Spearman correlations between bulk RNA-seq samples and their corresponding pseudobulk counterparts, maintaining identical proportions.

\subsection*{Analysis}

Despite the increasing popularity of using pseudobulk samples to train deep learning deconvolution models \citep{menden2020deep, chen2022deep}, no analyses have been conducted to compare the differences between bulk RNA-seq and pseudobulk samples. Here we evaluate the technical disparities between these two data types, seeking a better understanding of how these differences might impact data-driven models. To this end, we performed correlation (Pearson and Spearman) and differential expression (DE) analysis on four distinct datasets, where scRNA-seq/snRNA-seq and bulk RNA-seq (with known proportions) samples were available. For each dataset, we generated a pseudobulk matrix of the same size and proportions as the bulk RNA-seq samples, and repeated the analysis for several groups of top variant genes (100, 1k, 5k and 10k). See Methods for further details.

The first dataset was originated from human bone marrow tissue, encompassing both scRNA-seq \citep{ainciburu2023uncovering} and several bulk RNA-seq samples of HSC, MEP and GMP cells isolated via FACS. Our DE analysis between pseudobulk and bulk samples revealed that approximately 25\% of the genes exhibited differential expression, across both techniques (Supplementary Figure~\ref{fig:supp_DE_analysis}-A). In the correlation analysis, both Pearson and Spearman correlation coefficients exhibit a modest increase, reaching approximately 0.75 when a minimum of 1000 genes are taken into consideration (Figure~\ref{fig:ablation_study}A-C). These correlation results imply that the inclusion of at least 1000 genes might be preferable for achieving higher similarity between pseudobulk and bulk samples in this dataset.

Additionally, we explored two datasets derived from PBMC, consisting of a scRNA-seq matrix (\textit{data8k}, see Methods) and 2 microarray datasets (PBMC 1 and PBMC 2), with proportions determined by FACS. These showed about 25\% of the genes differentially expressed when considering a small number of them, but this percentage increases to half of the genes when a larger gene set is taken into account (Supplementary Figure~\ref{fig:supp_DE_analysis}-B). Correlation analyses conducted on both datasets demonstrate congruent outcomes with the Bone Marrow dataset, achieving an approximate Pearson correlation coefficient of 0.5 when at least 1000 genes are considered (Figure~\ref{fig:ablation_study}-D,E). While the Pearson correlation remains stable with an increasing number of genes, the Spearman correlation experiences a slight decrease. Similarly to the human Bone Marrow dataset, enhanced similarity between pseudobulk and bulk samples is achieved by considering a minimum of 1000 genes.

We extended our analysis to the human Brain Cortex dataset, incorporating snRNA-seq \citep{yang2022human} and bulk RNA-seq samples \citep{patrick2020deconvolving} with cell type proportions determined through IHC image analysis. The DE analysis highlights that a significant portion of genes are differentially expressed, and this percentage significantly increases as more genes are considered (Supplementary Figure~\ref{fig:supp_DE_analysis}-C). Notably, the correlation between pseudobulk and bulk samples in this dataset is comparatively low for both Pearson and Spearman correlations (Figure~\ref{fig:ablation_study}-F).

These findings emphasize significant distinctions between pseudobulk and actual bulk RNA-seq samples, particularly when only a low number of variant genes is available. However, the number of cells for building pseudobulk shows minimal variations with a sufficiently large gene selection. Additionally, they underscore differences between snRNA-seq and bulk RNA-seq, emphasizing the need for further study on complex tissues like the brain, where scRNA-seq is impractical due to the challenges in isolating specific cell types.

\newpage
\section{Interpretability analysis of DeepLIFT on Brain Cortex dataset}\label{notes:deeplift}

Regarding the human brain cortex dataset, many of these top genes were well-established canonical markers (Supplementary Figure \ref{fig:supp_deeplift_rosmap}-A). However, some, including PFKFB2 (astrocyte) \citep{muraleedharan2020ampk}, MAOB (astrocyte) \citep{nam2022revisiting}, SGPP2 (endothelial) \citep{venkataraman2008vascular}, and TF (oligodendrocyte) \citep{cheli2023transferrin}, were recently identified as enriched in their respective cell types. For instance, TF was newly recognized as a crucial gene for oligodendrocyte function and homeostasis, emphasizing that genes with lower expression levels yet high functional relevance also drive the deconvolution process.

Likewise, top terms found in our functional analysis of the Brain Cortex dataset also corresponded with canonical cell-functions (Supplementary Figure~\ref{fig:supp_deeplift_rosmap}-B). For instance, astrocytes play a key role in the brain function and metabolism as bridge cells between the blood vessels and neurons \citep{sofroniew2010astrocytes}. They regulate the flow of nutrients and neurotransmitter precursors necessary for the neurons, thus, it is not surprising that top enriched pathways reveal multiple transport processes like aspartate's, an amino-acid and neurotransmitter. Results on other cell types (except for the microglia with only three pathways) also support evidence that Sweetwater training process involves learning real and well-defined cell-specific biological functions.

%% file: supp_figures.tex
\setcounter{figure}{0} 
\renewcommand{\figurename}{Supplementary Figure}
\renewcommand{\thefigure}{S\arabic{figure}}


\begin{figure}[H]
\begin{center}
\includegraphics[width=\linewidth]{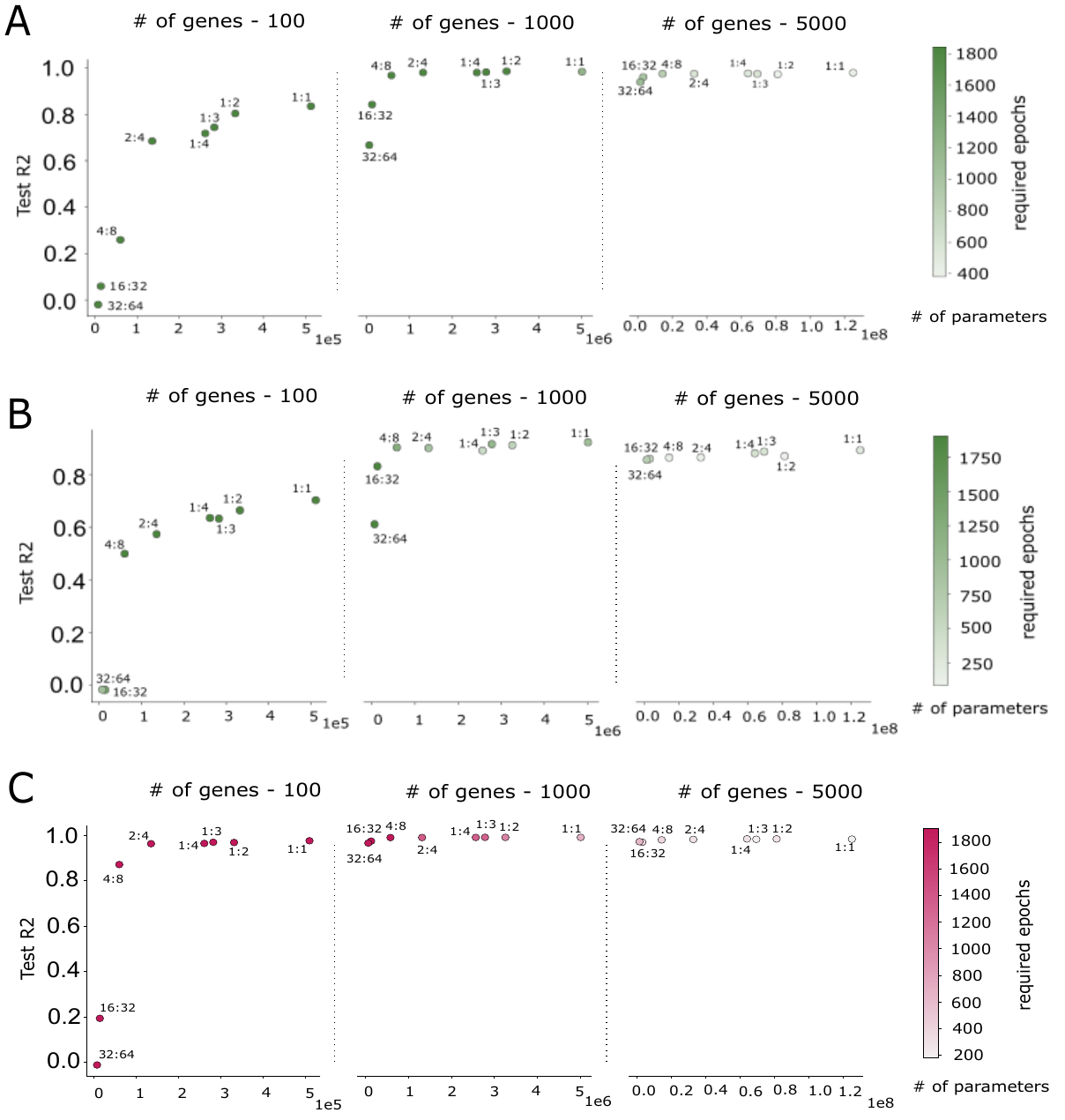}
\end{center}
\caption{\textbf{Architecture Analysis}. Evaluation on the performance acquired by Sweetwater on A) PBMC-1 B) PBMC-2 and C) Brain Cortex when used different number of input genes and different architectures. Concretely, different ratios were used in the autoencoder, which are reflected as text next to each point in the figure. For instance, (4,8) ratio would mean the encoder hidden layer has $\frac{1}{4}$ of the input neurons (genes), and the latent space has $\frac{1}{8}$ of the input neurons. The x-axis represent the number of trainable parameters each architecture has, the barplot shows the required epochs for the training until the EarlyStopping was triggered or the total number of epochs (30000/batch\_size) was achieved. The y-axis shows the $R^{2}$ between ground truth and predicted proportions on the test fold. It can be appreciated that, except for the case where a very low number of genes is available (first column only consider the top 100 most variant genes), our selected architecture (2,4), i.e., with ratio (1:$\frac{1}{2}$:$\frac{1}{4}$) achieves optimal test performance with the minimum number of epochs and parameters. For instance, when the number of variant genes is 5000, the architectures that obtain the best ratio of trainable parameters respect to performance are (32,64), (16,32), (4,8) and (2,4). Of those, (16,32) and (32,64) fail to obtain optimal performance when number of genes is reduced to 1000 most variant genes. Only (4,8) and (2,4) are able to do so, being (4,8) the one that is not able to deconvolve proportions efficiently when the number of genes is reduced to 100 on Brain Cortex dataset. }
\label{fig:ablation_supp_architecture}
\end{figure}

\newpage
\begin{figure}[H]
\begin{center}
\includegraphics[width=\linewidth]{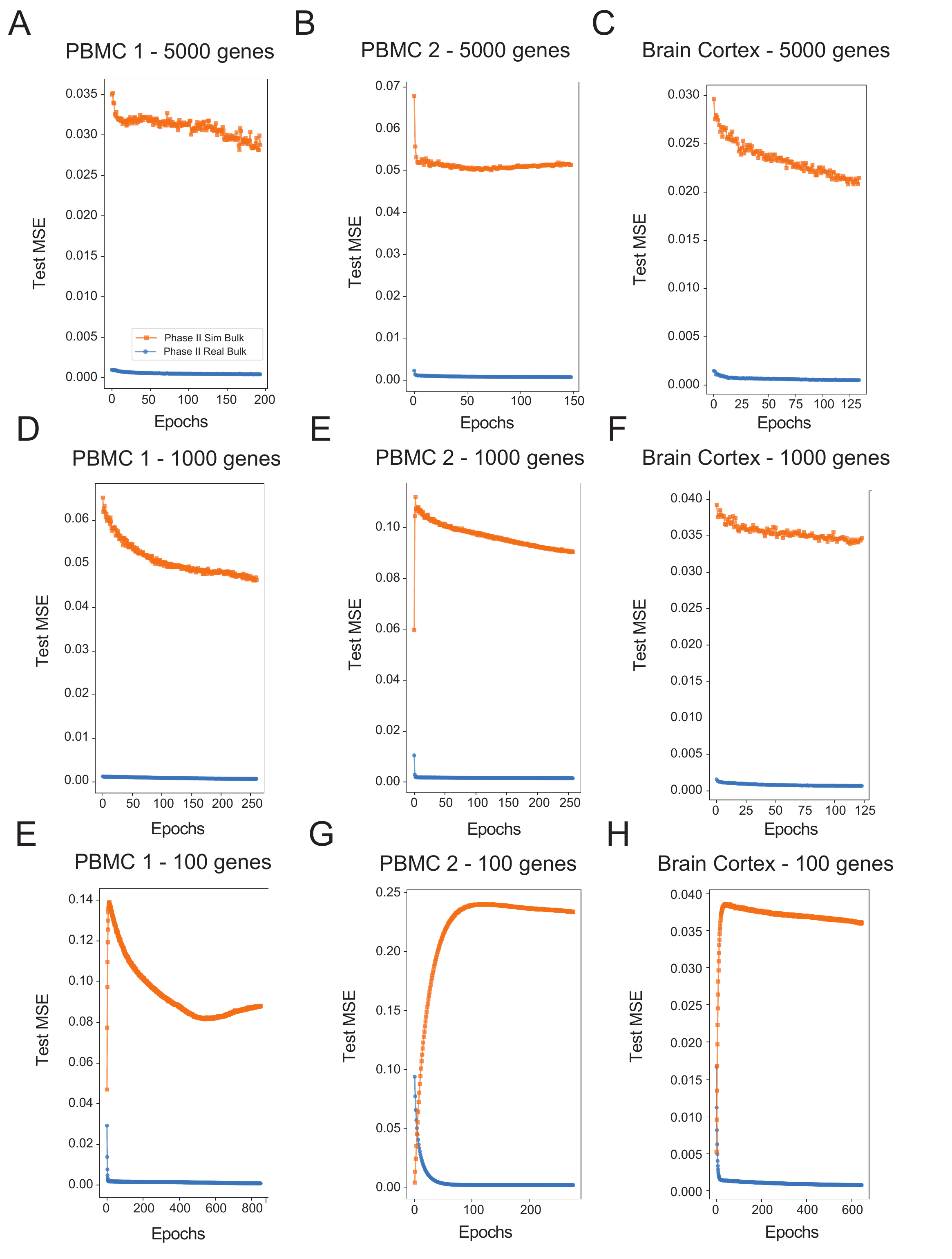}
\end{center}
\caption{\textbf{Loss function analysis on Phase II}. Evaluation of Test Mean Squared Error on real and simulated samples during Phase II training. It can be appreciated that as the number of selected genes increases, the model is able to better reconstruct both simulated and real samples, even though Phase I and Phase II are not jointly imposed. This suggests that the built latent space is representing common patterns among real and simulated samples, and by minimizing the reconstruction loss on the real ones, the simulated reconstruction is also being minimized.}
\label{fig:supp_losses_analysis}
\end{figure}

\begin{figure}[H]
\begin{center}
\includegraphics[width=0.5\linewidth]{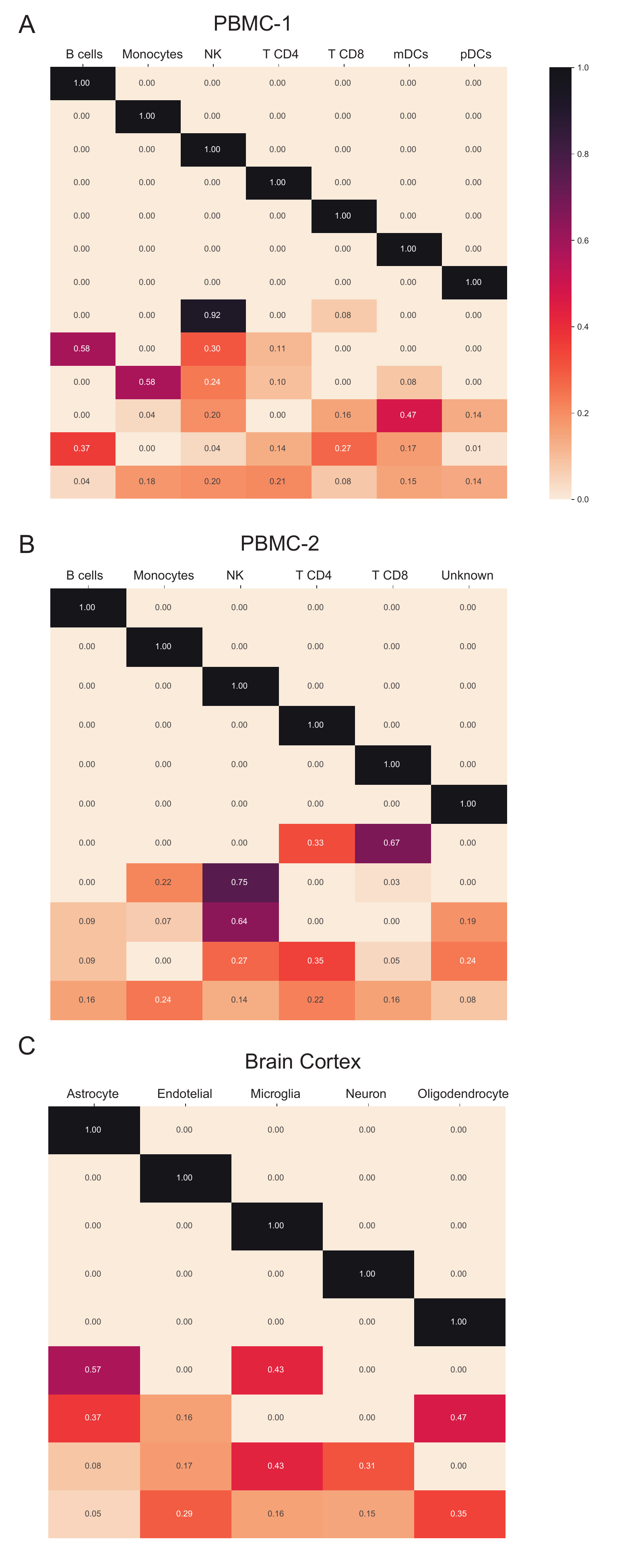}
\end{center}
\caption{\textbf{Test matrices generated for each of the evaluated datasets}. Ground truth proportions matrices generated for A) PBMC-1 (13 samples) B) PBMC-2 (11 samples) and C) Brain Cortex (9 samples) datasets. These matrices have been generated as a random combination of cell types, making sure isolated samples are present and there is at least one sample per combination, i.e., 1 sample containing just 2 cell types, 3 cell types, etc.. up to the number of cell types.}
\label{fig:supp_test_matrices}
\end{figure}

\newpage
\begin{figure}[H]
\begin{center}
\includegraphics[width=0.7\linewidth]{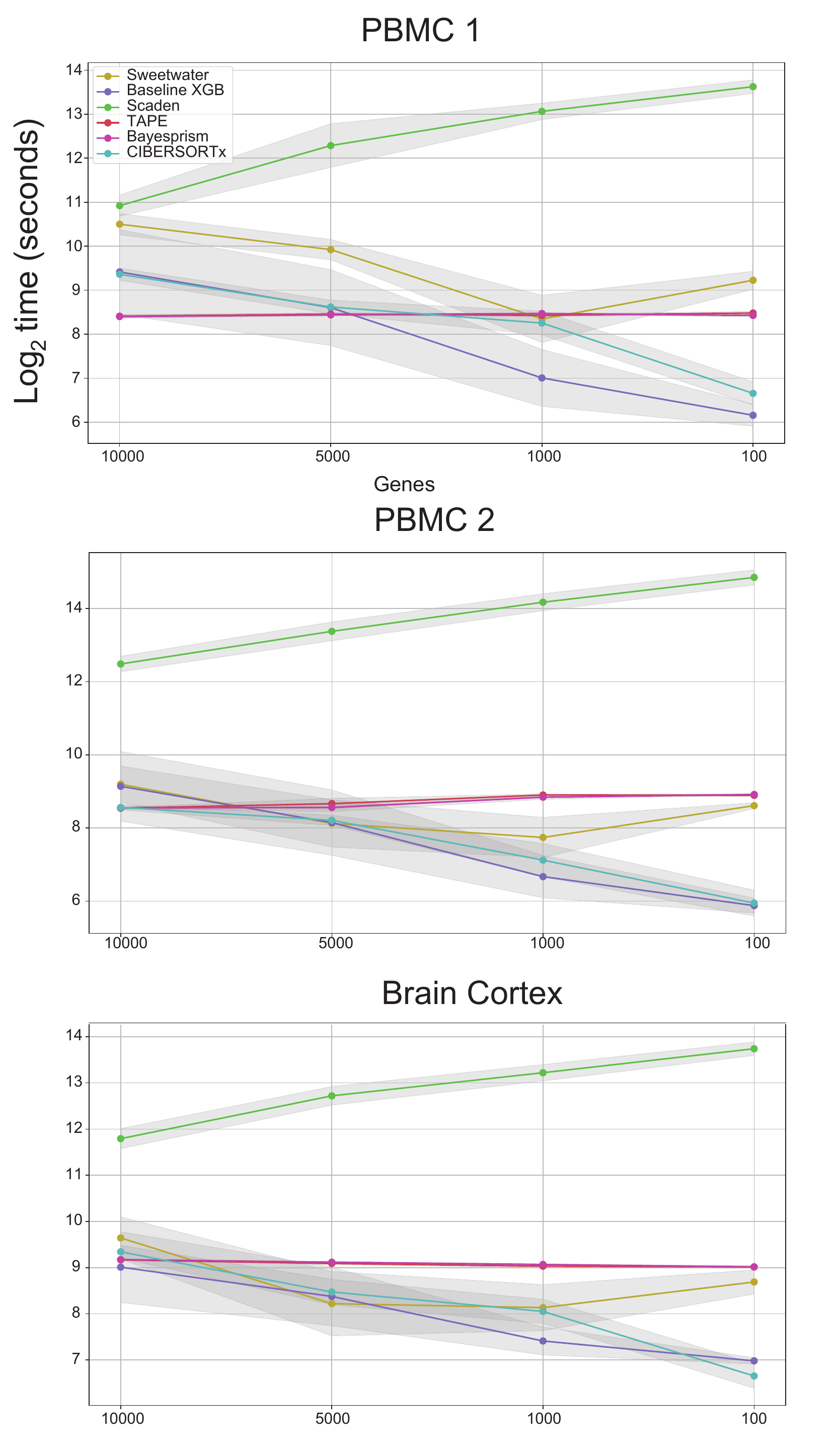}
\end{center}
\caption{\textbf{Deconvolution approaches benchmarking}. Time consumption in log2 seconds of all the evaluated methodologies for PBMC 1, PBMC 2 and Brain Cortex datasets.}
\label{fig:supp_time_comparison}
\end{figure}

\newpage
\begin{figure}[H]
\begin{center}
\includegraphics[width=\linewidth]{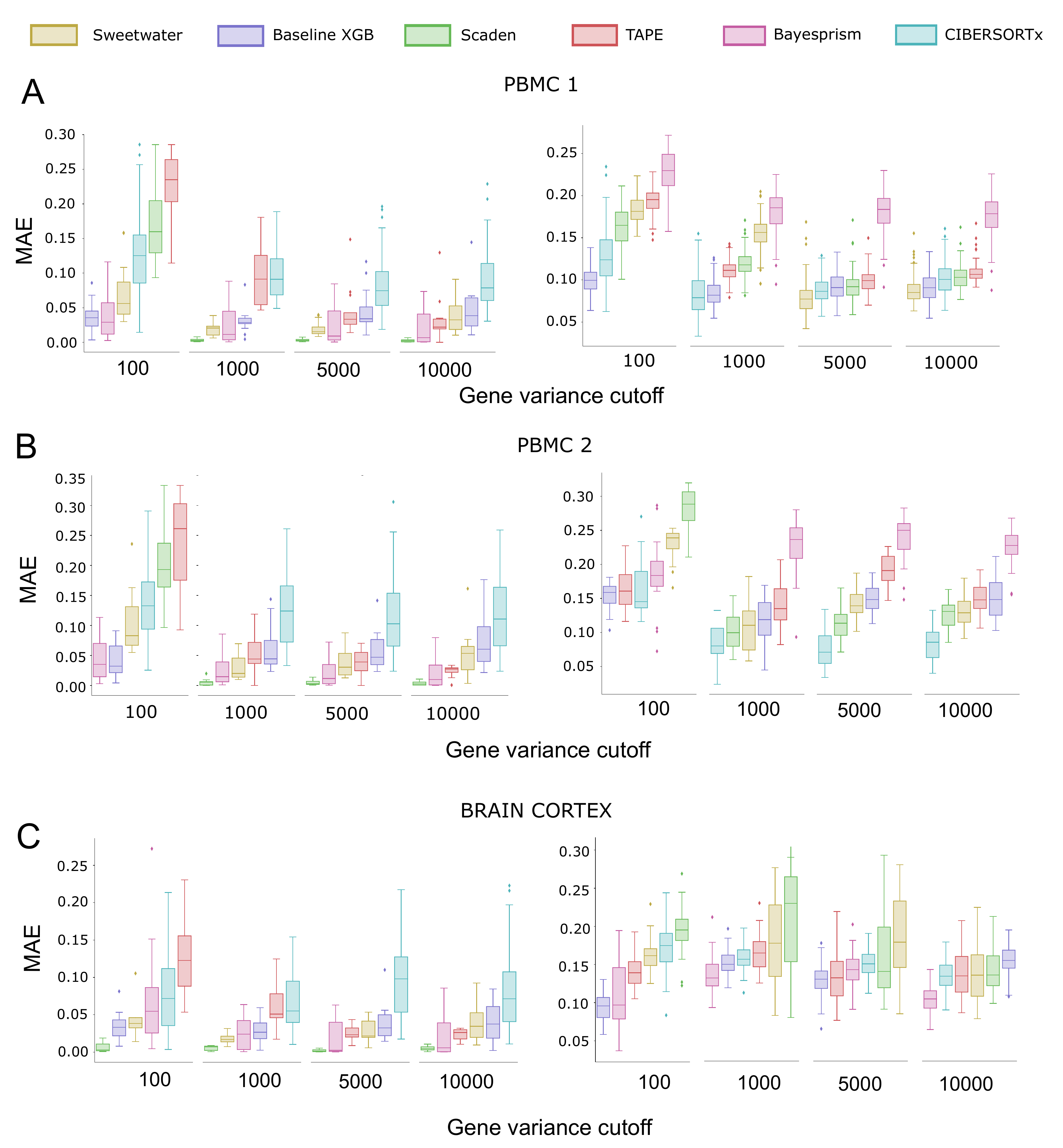}
\end{center}
\caption{\textbf{Deconvolution approaches benchmarking (MAE)}. Mean Absolute error (MAE) of deconvolved against ground truth proportions for PBMC-1 (\textbf{A}), PBMC-2 (\textbf{B}) and Brain Cortex dataset (\textbf{C}), for simulated (\textbf{right}) and realk bulk samples (\textbf{left}) accross different top variant gene cutoffs.}
\label{fig:supp_benchmarking_MAE}
\end{figure}

\newpage
\begin{figure}[H]
\begin{center}
\includegraphics[height=21cm]{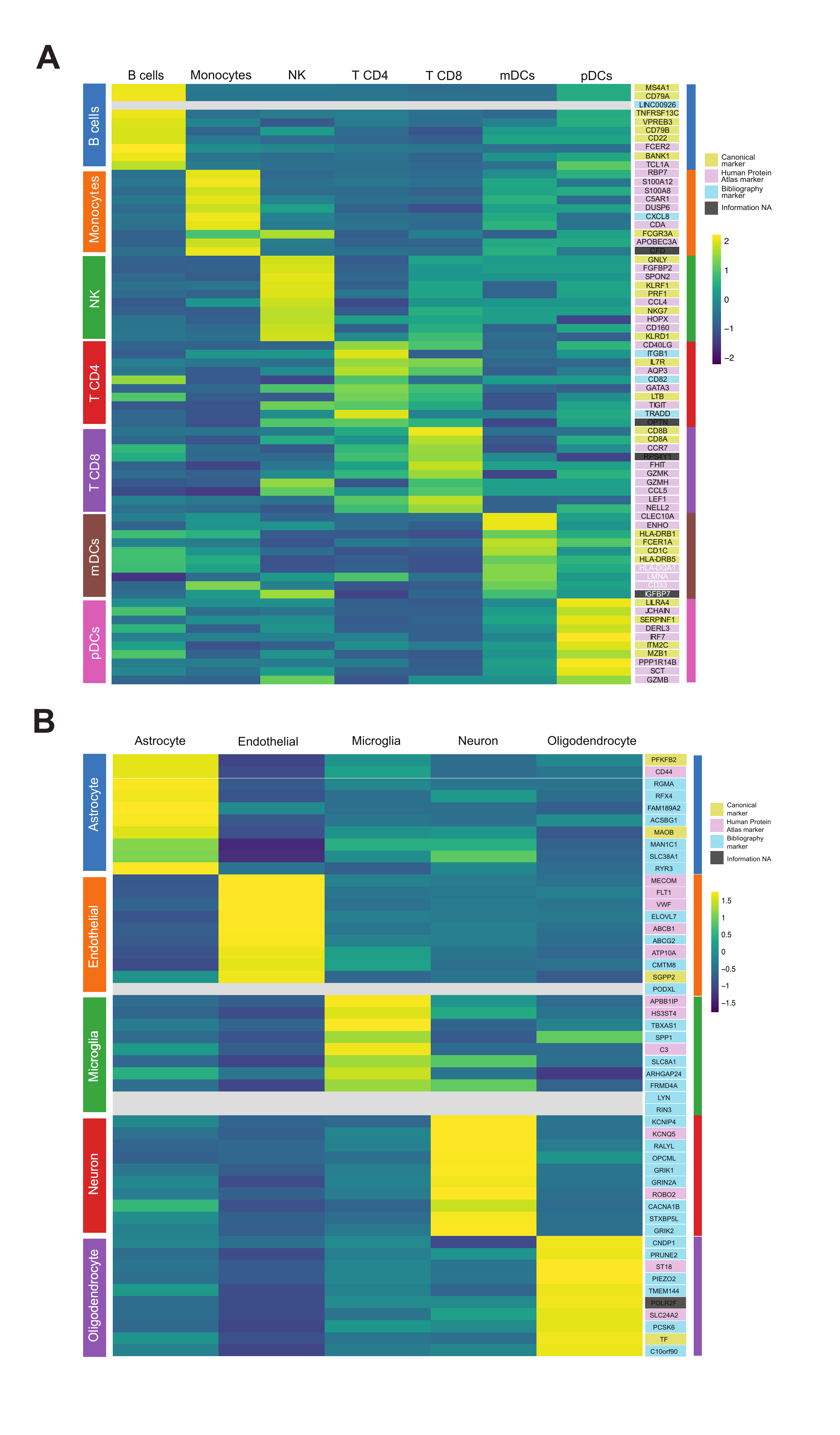}
\end{center}
\caption{\textbf{Differential expression analysis on reference scRNA-seq matrix}. Average log-fold change score for one cell type vs. the rest in the reference scRNA-seq matrix for A) PBMC-1 and B) Brain Cortex datasets.}
\label{fig:DE_deeplift}
\end{figure}

\newpage
\begin{figure}[H]
\begin{center}
\includegraphics[width=\linewidth]{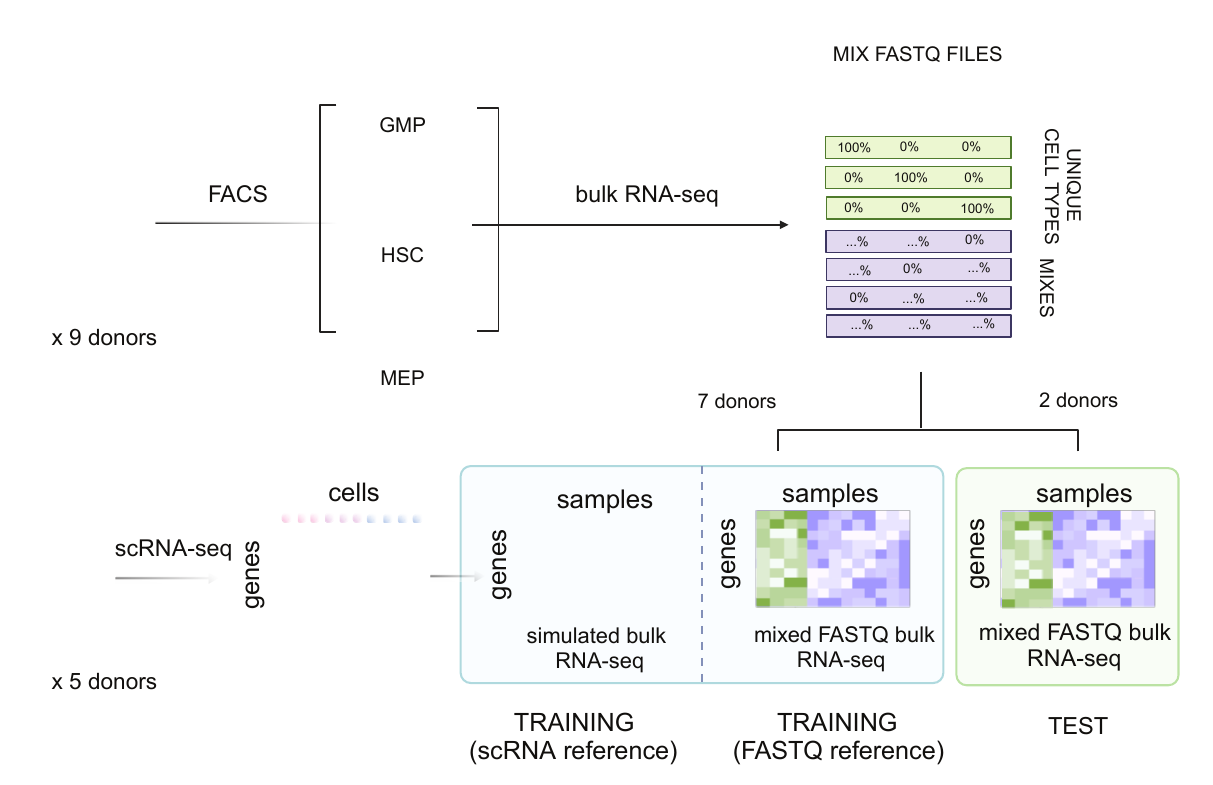}
\end{center}
\caption{Scheme depicting how Sweetwater is applied to deconvolve multiple simulated bulk RNA-seq samples coming from the mixing of two different donors' FASTQ files.}
\label{fig:supp_fastq}
\end{figure}

\newpage
\begin{figure}[H]
\begin{center}
\includegraphics[width=\linewidth]{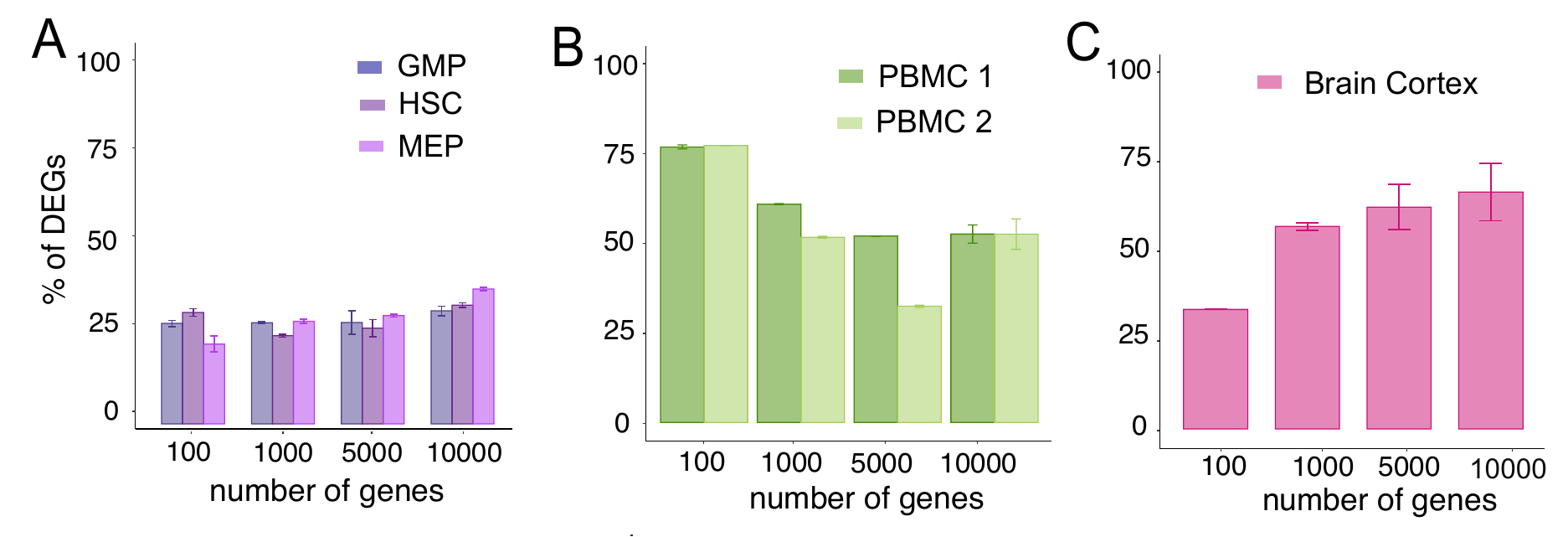}
\end{center}
\caption{\textbf{Differential expression and correlation analysis}. Percentage of differentially expressed genes (LFC $\ge$ 1 and p-value $\leq$ 0.05) for \textbf{A)} Bone marrow, \textbf{B)} PMBC and \textbf{C)} Brain Cortex datasets. }
\label{fig:supp_DE_analysis}
\end{figure}

\newpage
\begin{figure}[H]
\begin{center}
\includegraphics[height=21cm]{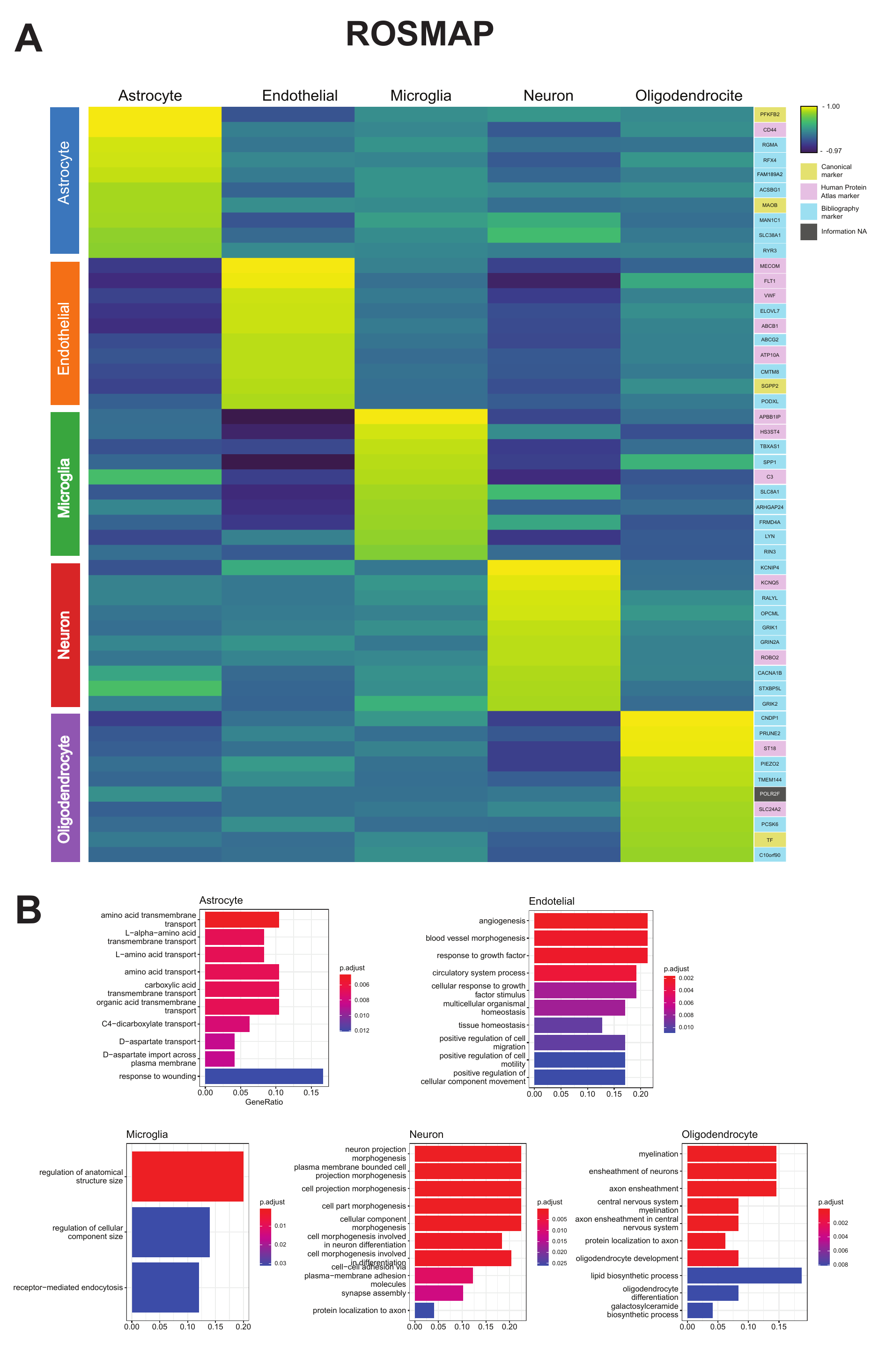}
\end{center}
\caption{\textbf{Sweetwater interpretation with DeepLIFT (Brain Cortex)}. A) Heatmap showing the top 10 genes for each cell type according to DeepLIFT score, when training Sweetwater on Brain Cortex dataset. B) Barplots showing top enriched GO terms (biological process) for the top 500 genes for each cell type according to DeepLIFT score, when training Sweetwater on Brain Cortex dataset.}
\label{fig:supp_deeplift_rosmap}
\end{figure}